\documentclass[preprint,12pt]{elsarticle}
\usepackage{amsmath}
\usepackage{amssymb}
\usepackage{amsfonts}
\usepackage{graphics}
\usepackage{graphicx}
\usepackage{amscd}
\usepackage{amsfonts}
\usepackage{epsf}
\usepackage{epsfig}
\usepackage{color}
\usepackage{feynmf}
\usepackage{wrapfig}
\usepackage{physics}
\usepackage{tikz}
\usepackage{ulem}
\usepackage{hyperref}

\usepackage{enumerate}
 \usepackage[utf8]{inputenc} 
 \usepackage{ascmac}

\biboptions{numbers,sort&compress}

\usepackage{etoolbox}
\makeatletter
\patchcmd{\ps@pprintTitle}
  {\let\@oddhead\@empty}
    {\def\@oddhead{\mbox{}\hfill \footnotesize{RUP-26-19}}}
  {}{}
\makeatother

\newcommand{\n}{\nonumber}

\newcommand{\eref}[1]{(\ref{#1})}

\allowdisplaybreaks[1]

\begin{document}

\title{Tensor Network Formulation of $\mathcal{PT}$-Symmetric Quantum Field Theory }
\author{Yasuyuki Hatsuda}
\address {Department of Physics, Rikkyo University, Toshima, Tokyo 171-8501, Japan}
\ead{yhatsuda@rikkyo.ac.jp}

\author{Kengo Kikuchi}
\address{Osaka Central Advanced Mathematical Institute (OCAMI), Osaka Metropolitan University, Osaka 558-8585, Japan}
\ead{kengo@yukawa.kyoto-u.ac.jp}

\date{\today}

\begin{abstract}
We develop a non-perturbative analytic tensor network formulation of $\mathcal{PT}$-symmetric scalar field theories defined on complex integration contours. Applying this formulation to the two-dimensional $\mathcal{PT}$-symmetric $\phi^4$ theory at negative quartic coupling, we derive an explicit analytic expression for the initial tensor and show that its components separate into even and odd sectors according to the parity of the sum of the tensor indices.
We further formulate the theory on an alternative complex contour and analytically establish, in arbitrary dimensions, an exact finite-volume relation between the lattice partition function defined on this contour and the real part of the analytic continuation of the Hermitian lattice $\phi^4$ partition function to negative quartic coupling.
\end{abstract}

\maketitle
\newpage
\tableofcontents
\newpage
\section{Introduction}

In conventional quantum theory, Hamiltonians are required to be Hermitian so that their spectra are real and the time evolution they generate is unitary. The pioneering work of Bender and Boettcher, however, showed that certain non-Hermitian Hamiltonians with unbroken $\mathcal{PT}$ symmetry can also possess entirely real spectra \cite{Bender:1998ke}. 
This discovery opened the way to a broader class of non-Hermitian quantum theories beyond the conventional Hermitian framework.
In particular, by imposing appropriate $\mathcal{PT}$-symmetric boundary conditions in the Hamiltonian formulation, or equivalently by choosing suitable complex integration contours in the path-integral formulation, one can define theories with non-Hermitian interactions or with potentials that are unbounded from below along the real axis \cite{Bender:2007nj}.

A variety of $\mathcal{PT}$-symmetric scalar field theories have attracted considerable attention.
Among theories with explicitly non-Hermitian interactions, the $\mathcal{PT}$-symmetric $i\phi^3$ theory provides a field-theoretic description of the Yang--Lee edge singularity \cite{Fisher:1978pf}. It has also been investigated in the context of renormalization-group behavior and critical phenomena \cite{Bender:2012ea,Bender:2013qp}. More recently, the relation between multicritical $\mathcal{PT}$-symmetric Ginzburg--Landau theories and non-unitary minimal models has been actively investigated \cite{Lencses:2024wib,Katsevich:2025ojk}.

Another important class consists of $\mathcal{PT}$-symmetric theories with real potentials that are unbounded from below along the real field axis. A paradigmatic example is the $\mathcal{PT}$-symmetric $-g\phi^4$ theory with $g>0$, which was investigated using Schwinger--Dyson and bound-state methods \cite{Bender:1999ek, Bender:2001qx}. More recent studies have examined this theory using complex integration contours and explored its relation to the analytic continuation of the Hermitian $\lambda\phi^4$ theory to negative quartic coupling \cite{Ai:2022csx,Lawrence:2023woz}.

Tensor network methods have also been developed as non-perturbative tools for lattice field theories and applied to both real and complex
scalar $\phi^4$ models \cite{Kadoh:2019ube, Delcamp:2020hzo, Meurice:2020pxc}.


In this paper, we develop a non-perturbative analytic tensor network formulation of $\mathcal{PT}$-symmetric scalar field theories defined by complex integration contours. We first represent path integrals along a class of complex contours in terms of tensor networks and then specialize the formulation to the $\mathcal{PT}$-symmetric $-g\phi^4$ theory. 
Although its potential is real along the real field axis, it is unbounded from below there, and hence the conventional path integral along the real contour diverges. Nevertheless, the theory can be consistently defined by choosing appropriate complex integration contours compatible with $\mathcal{PT}$ symmetry.


We obtain two main results. First, applying the formulation to the two-dimensional $-g\phi^4$ theory on a cone-shaped contour, we derive an explicit analytic expression for the initial tensor and show that its components exhibit a characteristic non-perturbative structure. In particular, the tensor components separate into even and odd sectors according to the parity of the sum of the tensor indices: the even sector components are real and admit a conventional perturbative expansion, whereas the odd sector components are purely imaginary and encode exponentially small non-perturbative contributions. Furthermore, for each fixed set of indices in the odd sector, the non-perturbative contribution is expressed exactly as an exponential factor multiplied by a finite polynomial.
Second, for the theory defined on a wedge-type contour, we analytically establish, in arbitrary dimensions, an exact finite-volume relation between the wedge-contour partition function and the real part of the analytic continuation of the Hermitian $\lambda\phi^4$ lattice partition function to negative quartic coupling.
An important feature of our formulation is that the tensor network method is used not merely as a numerical framework for coarse graining and evaluating partition functions \cite{Levin:2006jai}, but also as an analytic framework for extracting non-perturbative information directly from local tensor components.

The remainder of this paper is organized as follows. In Sec.~2, we formulate path integrals over a class of complex integration contours in a tensor network representation. 
In Sec.~3, applying this formulation to the $\mathcal{PT}$-symmetric $-g\phi^4$ theory on a cone contour, we derive an explicit analytic expression for the corresponding initial tensor and analyze its parity structure. 
In Sec.~4, we formulate the theory on a wedge contour, establish its relation to the analytically continued Hermitian $\lambda\phi^4$ theory, and extend this relation to arbitrary dimensions. Section~5 is devoted to a summary and discussion.
Appendix A defines the tensor trace notation. Appendix B summarizes the
confluent hypergeometric functions and related formulae used in the main
text. Appendix C presents numerical tests of the tensor network formulation
in quantum mechanics.

\section{Tensor network representation on complex contours}


We consider scalar field theories whose path integrals are defined by extending the integration contour from the real field axis to a contour in the complex field plane. After discretizing the theory on a lattice, we apply a tensor network construction to the resulting lattice partition function. In this section, we start from the standard construction based on the Taylor expansion of the exponential of the nearest-neighbor interaction, originally developed for lattice $\phi^4$ scalar field theories with integration over the real field axis \cite{Delcamp:2020hzo}, and adapt it to theories defined on a complex contour.

For a lattice theory with a local potential $V(\phi)$, this expansion factorizes the Boltzmann factor associated with each nearest-neighbor interaction and allows the local initial tensor to be expressed as a one-variable contour integral. This feature becomes important once a specific contour is chosen, since the contour dependence is encoded directly in the local tensor. The tensor network representation is therefore particularly well suited to $\mathcal{PT}$-symmetric theories defined on complex integration contours.

\subsection{Setup and lattice discretization}

Consider a two-dimensional Euclidean scalar field theory with a general local potential $V(\phi)$. In a  continuum theory, the action is given by
\begin{align}
S[\phi]=\int d^2x
\left[
\frac{1}{2}(\partial_\mu\phi)^2
+
V(\phi)
\right].
\end{align}
The $\mathcal{PT}$ transformation is defined by
\begin{align}
\mathcal{P}:\phi\rightarrow-\phi,
\qquad
\mathcal{T}:\phi\rightarrow\phi,
\quad
i\rightarrow-i .
\end{align}
The local potential and the integration contour are assumed to be chosen so that the path-integral formulation is compatible with this $\mathcal{PT}$ transformation.
The path integral is defined by integrating the field variable at each spacetime point along a contour $\mathcal{C}$ in the complex $\phi$ plane:
\begin{align}
Z
=\int_{\phi(x)\in\mathcal{C}}
\mathcal{D}\phi \,e^{-S[\phi]}.
\end{align}

To construct a tensor network representation, the theory is discretized on a finite two-dimensional $N\times N$ square lattice $\Lambda$  with periodic boundary conditions. We denote the lattice volume by $V=N^2$. Hereafter, we work in lattice units and set the lattice spacing $\epsilon$ to unity. 
The lattice partition function is
\begin{align}
Z
=
\prod_{i\in\Lambda}
\int_{\mathcal{C}} d\phi_i\,
e^{-S_{\mathrm{lat}}[\phi]},
\end{align}
with
\begin{align}
S_{\mathrm{lat}}[\phi]&=
\sum_{\langle i,j\rangle}
\frac{(\phi_i-\phi_j)^2}{2}
+
\sum_{i\in\Lambda}
V(\phi_i)\\
&=\sum_{\langle i,j\rangle}
\left(\frac{\phi_i^2}{2}+\frac{\phi_j^2}{2}-\phi_i\phi_j\right)
+
\frac{1}{4}\sum_{\langle i,j\rangle}
\left\{V(\phi_i)+V(\phi_j) \right\}.
\end{align}
Here $\langle i,j\rangle$ denotes a pair of nearest-neighbor sites.

\subsection{Local tensor on a complex contour}
To formulate the integral on the contour, we decouple the variables and reduce the integration to a one-variable integral. We then Taylor-expand the cross term as follows \cite{Delcamp:2020hzo};
\begin{align}
\exp\left[ \phi_i \phi_j \right] = \sum_{n=0}^\infty \frac{\phi^n_i \phi_j^n}{n!}.
\end{align}
We introduce the one-leg factor as
\begin{align}
f_l(\phi)=\frac{\phi^l}{\sqrt{l!}}\exp\left[ -\frac{\phi^2}{2}-\frac{1}{4}V(\phi) \right].
\end{align}
Using this one-leg factor, we define the initial tensor by the one-dimensional integral form:
\begin{align}
T_{abcd}&=\int_{\mathcal C} d\phi\, f_a(\phi)f_b(\phi)f_c(\phi)f_d(\phi)\\
&=\int_{\mathcal C}d\phi \frac{\phi^{a+b+c+d}}{\sqrt{a!b!c!d!}}\exp\left[ -2\phi^2-V(\phi) \right]\\
&=\frac{h_{\mathcal{C}}(j)}{\sqrt{a!b!c!d!}},
\end{align}
where $j=a+b+c+d$ and
\begin{align}
h_{\mathcal{C}}(j)\equiv \int_{\mathcal C} d\phi\, \phi^{j}\exp\left[ -2\phi^2-V(\phi) \right].
\end{align}
The integration contour $\mathcal{C}$ must be chosen to be invariant under $\mathcal{PT}$ symmetry. 
One possible choice, which we consider in the following section, is the cone-shaped contour defined in \eqref{cone}.
Finally, the partition function in the tensor network formalism is obtained as
\begin{align}
Z=\mathrm{tTr}\left[ T_{abcd}\right].
\end{align}
Here, the notation $\mathrm{tTr}$ denotes the tensor trace defined in \ref{tensortrace}.

\subsection{Extension to arbitrary dimensions}

At the level of the local tensor representation, this construction can be extended straightforwardly to arbitrary dimensions. Although the efficient contraction or coarse graining of higher-dimensional tensor networks is a separate and difficult problem, it will not be addressed in this work. This observation is nevertheless useful for our purposes, since the analytic structure of the local tensor, rather than the contraction algorithm itself, is the main object of the discussion below. 

On a $d$-dimensional hypercubic lattice, each site has $2d$ nearest-neighbor links. 
Accordingly, the local tensor has $2d$ indices. 
The lattice action can be rewritten as
\begin{align}
S_{\mathrm{lat}}[\phi]
=
\sum_{\langle i,j\rangle}
\left[
\frac{\phi_i^2}{2}
+
\frac{\phi_j^2}{2}
-
\phi_i\phi_j
+
\frac{1}{2d}V(\phi_i)
+
\frac{1}{2d}V(\phi_j)
\right].
\end{align}
We then introduce the one-leg factor
\begin{align}
f_l(\phi)
=
\frac{\phi^l}{\sqrt{l!}}
\exp\left[
-\frac{\phi^2}{2}
-
\frac{1}{2d}V(\phi)
\right].
\end{align}
The corresponding initial tensor is given by
\begin{align}
T_{n_1 n_2\cdots n_{2d-1} n_{2d}}
&=
\int_{\mathcal C}d\phi\,
f_{n_1}(\phi)f_{n_2}(\phi)
\cdots
f_{n_{2d-1}}(\phi)f_{n_{2d}}(\phi)\\
&=\int_{\mathcal C}d\phi \left[\prod_{i=1}^{2d}\frac{\phi^{n_i}}{\sqrt{n_i !}}\right]\exp\left[ -d\phi^2-V(\phi) \right]\\
&=\frac{h^d_{\mathcal{C}}(j)}{\prod_{i=1}^{2d}\sqrt{n_i !}},
\end{align}
where $j=\sum_{i=1}^{2d}n_{i}$ and 
\begin{align}
h^d_{\mathcal{C}}(j)\equiv \int_{\mathcal C} d\phi\, \phi^{j}\exp\left[ -d\phi^2-V(\phi) \right].
\end{align}
The partition function in $d$ dimensions is then given by
\begin{align}
Z^{\mathcal{PT}}_{d}=\mathrm{tTr}\left[ T_{n_1 n_2\cdots n_{2d-1} n_{2d}}\right].
\end{align}

In the following section, the contour formulation developed above is specialized to a $\mathcal{PT}$-symmetric theory with an unbounded potential.

\section{Cone-contour formulation of the $\mathcal{PT}$-symmetric $-g\phi^4$ theory}

We now apply the tensor network formulation developed above to a concrete $\mathcal{PT}$-symmetric scalar field theory. We consider the $-g\phi^4$ theory, whose potential is unbounded from below along the real field axis. Consequently, the path integral along the real contour diverges, and the theory must instead be defined using suitable integration contours in the complex $\phi$ plane.

Here we choose a cone-shaped contour that provides a convergent $\mathcal{PT}$-symmetric definition of the theory. The two asymptotic rays of the contour are mapped into each other under the $\mathcal{PT}$ transformation, and the integrand is exponentially suppressed at large field amplitude. Although other contours may also be compatible with $\mathcal{PT}$ symmetry, the cone contour is particularly convenient for the analytic tensor network construction developed below.
A direct evaluation of the lattice path integral along this contour is cumbersome. 
At each lattice site, the field variable is integrated along either of the two rays, so that a lattice with $V$ sites gives rise to $2^V$ possible assignments of integration rays.
The tensor network representation avoids this exponential proliferation. By expanding the exponential of the nearest-neighbor interaction, the field integrations factorize site by site, and all dependence on the contour is encoded in a single one-variable integral defining the local tensor. This reduction makes it possible to analyze the local tensor explicitly and to identify its characteristic perturbative and non-perturbative structures.


\subsection{The $\mathcal{PT}$-symmetric $-g\phi^4$ theory}

We consider the two-dimensional Euclidean scalar field theory with action
\begin{align}
S[\phi]
=
\int d^2x
\left[
\frac{1}{2}(\partial_\mu\phi)^2
+
\frac{m_0^2}{2}\phi^2
-
\frac{g_{\mathrm{cont}}}{4}\phi^4
\right],
\end{align}
where $g_{\mathrm{cont}}>0$. Along the real axis, the potential is unbounded below because of the negative quartic term, and the path integral is not defined as an ordinary real-contour integral. 
The dimensionless bare lattice parameters used below are related to the continuum parameters by
\begin{align}
\mu_0^2
=
\epsilon^2m_0^2,
\qquad
g
=
\epsilon^2g_{\mathrm{cont}}>0.
\end{align}
As in Sec.~2, we set $\epsilon=1$ in the following.
After discretization on a two-dimensional square lattice, the lattice action becomes
\begin{align}
S_{\mathrm{lat}}[\phi]&=
\sum_{\langle i,j\rangle}
\frac{(\phi_i-\phi_j)^2}{2}
+
\sum_{i\in\Lambda}
\left[
\frac{\mu_0^2}{2}\phi_i^2
-
\frac{g}{4}\phi_i^4
\right].\label{lateq}
\end{align}

In this work, we define the $\mathcal{PT}$-symmetric theory by choosing the cone contour in the complex $\phi$ plane,
\begin{align}
\mathcal C^{\mathrm{cone}}:
\phi=s\left(
e^{\frac{\pi i}{4}}\theta(-s)+e^{-\frac{\pi i}{4}}\theta(s)\right),\qquad s\in\mathbb R .\label{cone}
\end{align}
This contour is compatible with $\mathcal{PT}$ symmetry, since its two asymptotic rays are mapped into each other under the $\mathcal{PT}$ transformation. It also ensures that the quartic part of the exponential weight is exponentially suppressed: along the two rays one has $\phi^4=-s^4$, so that the wrong-sign quartic term gives a damping factor $\exp(-g s^4/4)$ at large $|s|$.

\subsection{Initial tensor on the cone contour}

We now construct the initial tensor for the theory defined on the cone contour.
 A direct evaluation of the cone-contour path integral is difficult, since the field variable at each lattice site can be integrated independently along either of the two rays of the contour.
 For a lattice with $V$ sites, the integration domain therefore decomposes into $2^{V}$ ray sectors, making a direct evaluation of the many-variable path integral impractical.

The tensor network formalism avoids this difficulty: after the Taylor expansion of the nearest-neighbor coupling, the field integrations factorize site by site, and the contour dependence is reduced to a single local one-variable integral defining the initial tensor. 
This is the key reason why the tensor network representation is naturally adapted to the cone-contour definition of the $\mathcal{PT}$-symmetric theory.

As shown in the previous section, the Taylor expansion of the nearest-neighbor coupling leads to the one-leg factor
\begin{align}
f_l(\phi)=\frac{\phi^l}{\sqrt{l!}}\exp\left[ -\left(\frac{1}{2}+\frac{\mu_0^2}{8} \right)\phi^2+\frac{g}{16}\phi^4 \right].
\end{align}
Using this one-leg factor, we define the initial tensor as
\begin{align}
T^{\mathrm{cone}}_{abcd}&=\int_{\mathcal C^{\mathrm{cone}}} d\phi\, f_a(\phi)f_b(\phi)f_c(\phi)f_d(\phi)\\
&=\int_{\mathcal C^{\mathrm{cone}}}d\phi\, \frac{\phi^{a+b+c+d}}{\sqrt{a!b!c!d!}}\exp\left[ -A\phi^2+\frac{g}{4}\phi^4 \right]\\
&=\frac{h^{\mathrm{cone}}(j, A, g)}{\sqrt{a!b!c!d!}},
\end{align}
where $j\equiv a+b+c+d$, $A\equiv 2+\mu_0^2/2>0$, and
\begin{align}
h^{\mathrm{cone}}(j, A, g)\equiv \int_{\mathcal C^{\mathrm{cone}}} d\phi\, \phi^{j}
\exp\left[ -A\phi^2+\frac{g}{4}\phi^4 \right].
\end{align}
The effect of the cone contour is entirely contained in this local one-variable integral.
Using the parametrization of the cone contour in \eref{cone}, the local integral can be decomposed into the contributions from the two rays:
\begin{align}
h^{\mathrm{cone}}(j, A, g)&=\int_{-\infty}^{0}ds e^{\frac{\pi i}{4}}
\left(s e^{\frac{\pi i}{4}}\right)^j
\exp\left[-iAs^2-\frac{g}{4}s^4\right]\nonumber\\
&\quad
+\int_{0}^{\infty}ds
e^{-\frac{\pi i}{4}}
\left(s e^{-\frac{\pi i}{4}}\right)^j
\exp\left[+iAs^2-\frac{g}{4}s^4\right].
\end{align}
Changing the integration variable $s\to -s$ in the first integral, this becomes
\begin{align}
h^{\mathrm{cone}}(j, A, g)
=
(-1)^j
e^{\frac{\pi i}{4}(1+j)}
I_j(A,g)
+
e^{-\frac{\pi i}{4}(1+j)}
I_j^*(A,g),
\end{align}
where
\begin{align}
I_j(A,g)
&\equiv
\int_0^\infty ds\,
s^j
\exp\left[
-iAs^2
-
\frac{g}{4}s^4
\right],
\\
I_j^*(A,g)
&\equiv
\int_0^\infty ds\,
s^j
\exp\left[
+iAs^2
-
\frac{g}{4}s^4
\right].
\end{align}
Here $I_j^*(A,g)$ is the complex conjugate of $I_j(A,g)$ for real $A$ and $g>0$.
The integral $I_j(A,g)$ can be expressed in terms of the Tricomi function as
\begin{align}
I_j(A,g)
=
\frac{1}{2}
\left(\frac{1}{g}\right)^{\frac{1+j}{4}}
\Gamma\left(\frac{1+j}{2}\right)
U\left(
\frac{1+j}{4},
\frac{1}{2},
z_+
\right),
\end{align}
and similarly
\begin{align}
I_j^*(A,g)
=
\frac{1}{2}
\left(\frac{1}{g}\right)^{\frac{1+j}{4}}
\Gamma\left(\frac{1+j}{2}\right)
U\left(
\frac{1+j}{4},
\frac{1}{2},
z_-
\right).
\end{align}
Here we have introduced
\begin{align}
z_\pm &\equiv X e^{\pm i (\pi-0)}= -X\pm i0, \quad X \equiv \frac{A^2}{g}.
\end{align}
The subscripts $+$ and $-$ label the two sides of the branch cut along the negative real axis, corresponding respectively to approaching the cut from above, $+i0$, and from below, $-i0$.

Therefore the cone-contour initial tensor is
\begin{align}
T^{\mathrm{cone}}_{abcd}
=
\frac{1}{2}
\frac{1}{\sqrt{a!b!c!d!}}
\left(\frac{1}{g}\right)^{\frac{1+j}{4}}
\Gamma\left(\frac{1+j}{2}\right)
\left[
(-1)^j
e^{\frac{\pi i}{4}(1+j)}
U_+
+
e^{-\frac{\pi i}{4}(1+j)}
U_-
\right],
\end{align}
where
\begin{align}
U_\pm
\equiv
U\left(
\frac{1+j}{4},
\frac{1}{2},
z_\pm
\right).
\end{align}
Then the cone-contour partition function is therefore given by
\begin{align}
Z^{\mathrm{cone}}
=
\mathrm{tTr}\left[
T_{abcd}^{\mathrm{cone}}
\right].
\end{align}
Since the relative sign between the two ray contributions depends on the parity of $j$, we now analyze the even and odd sectors separately.

\subsection{Even and odd sectors}

For the parity analysis, it is useful to rewrite the exact expression for the cone-contour tensor in terms of the two boundary values of the Tricomi function. We define
\begin{align}
\tilde{T}_{\pm,abcd}
&=e^{\mp i\pi \frac{1+j}{4}}\frac{1}{\sqrt{a!b!c!d!}}
\left(\frac{1}{g}\right)^{\frac{1+j}{4}}
\Gamma\left(\frac{1+j}{2}\right)
U\left(
\frac{1+j}{4},
\frac{1}{2},
z_{\mp}
\right).\label{39}
\end{align}
Then the cone-contour initial tensor can be written as
\begin{align}
T^{\mathrm{cone}}_{abcd}
=
\frac{1}{2}
\left[
\tilde{T}_{+,abcd}
+
(-1)^j
\tilde{T}_{-,abcd}
\right].\label{40}
\end{align}
This form makes the dependence on the parity of $j$ explicit. For even $j$, the tensor is given by the average of the values of the Tricomi function taken just above and below the branch cut, whereas for odd $j$, it is given by half of the discontinuity across the cut.


\subsubsection{The case of even $j$}

When $j$ is even, one has $(-1)^j=1$. The cone-contour tensor then becomes
\begin{align}
T_{abcd}^{\mathrm{cone}}
&=
\frac{1}{2}
\left(
\tilde{T}_{+,abcd}
+
\tilde{T}_{-,abcd}
\right)
\nonumber\\
&=
\frac{1}{2}
\frac{1}{\sqrt{a!b!c!d!}}
\left(\frac{1}{g}\right)^{\frac{1+j}{4}}
\Gamma\left(\frac{1+j}{2}\right)
\Sigma(X),
\end{align}
where
\begin{align}
\Sigma(X)
\equiv
e^{-i\pi \frac{1+j}{4}}U_-
+
e^{i\pi \frac{1+j}{4}}U_+ .
\end{align}
Since $\tilde{T}_{-,abcd}$ is the complex conjugate of $\tilde{T}_{+,abcd}$, the even-sector tensor components are real.

Using the connection formula for the Tricomi function, $\Sigma(X)$ can be written in terms of the Kummer function as
\begin{align}
\Sigma(X)
&=
2\sqrt{\pi}
\Bigg[
\frac{
\cos\left(
\frac{1+j}{4}\pi
\right)
}{
\Gamma\left(
\frac{3+j}{4}
\right)
}
M\left(
\frac{1+j}{4},
\frac{1}{2},
-X
\right)
\nonumber\\
&\quad
+
\frac{
2X^{\frac{1}{2}}
\sin\left(
\frac{1+j}{4}\pi
\right)
}{
\Gamma\left(
\frac{1+j}{4}
\right)
}
M\left(
\frac{3+j}{4},
\frac{3}{2},
-X
\right)
\Bigg].
\end{align}

We next consider the large-$X$ asymptotic expansion of $\Sigma(X)$, with $X=A^2/g$. From the expression above, one obtains
\begin{align}
\Sigma(X)
&\sim
2\left(\frac{g}{A^2}\right)^{\frac{1+j}{4}}
\sum_{n=0}^\infty \frac{(\frac{j+1}{4})_n (\frac{j+3}{4})_n}{n!} \biggl( \frac{g}{A^2} \biggr)^n
\end{align}
Clearly, this perturbative expansion is a divergent series, and one needs the Borel resummation technique to reconstruct the analytic function. Since $\Sigma(X)$ is real for $X>0$,
the standard prescription is the median resummation, in which the exponentially small imaginary ambiguities cancel. Thus, the even sector is described by the median resummation of the perturbative sector at the level of each local tensor component.

Substituting this expansion into the tensor, we find
\begin{align}
T_{abcd}^{\mathrm{cone}}
&\sim
\frac{1}{\sqrt{a!b!c!d!}}
\left(\frac{1}{A}\right)^{\frac{1+j}{2}}
\Gamma\left(\frac{1+j}{2}\right)\sum_{n=0}^\infty \frac{(\frac{j+1}{4})_n (\frac{j+3}{4})_n}{n!} \biggl( \frac{g}{A^2} \biggr)^n.
\end{align}
Therefore, for even $j$, the cone-contour initial tensor is described, under the median prescription, by an ordinary perturbative asymptotic expansion in powers of $g$ around $g=0$. In this sense, no explicit non-perturbative contribution appears in the even sector.


\subsubsection{The case of odd $j$}
In contrast, when $j$ is odd, one has $(-1)^j=-1$. The relative sign between the two terms is then reversed, and the cone-contour tensor becomes
\begin{align}
T_{abcd}^{\mathrm{cone}}
&=\frac{1}{2}\left(\tilde{T}_{+, abcd} -\tilde{T}_{-, abcd} \right)\\
&=\frac{1}{2}\frac{1}{\sqrt{a!b!c!d!}}\left(\frac{1}{g}\right)^{\frac{1+j}{4}}\Gamma\left(\frac{1+j}{2} \right)\Delta(X)
\end{align}
where
\begin{align}
\Delta(X)&\equiv e^{-i\pi \frac{1+j}{4}}U_- - e^{i\pi \frac{1+j}{4}}U_+
\label{eq:Delta-U}
\end{align}
Since $\tilde{T}_{-,abcd}$ is the complex conjugate of $\tilde{T}_{+,abcd}$, the odd-sector tensor components are purely imaginary.
Using the connection formula for the Tricomi function, $\Delta(X)$ can be written in terms of the Kummer function as
\begin{align}
\Delta(X)&=-2i \sqrt{\pi}\Bigg[\frac{\sin\left(\frac{1+j}{4}\pi  \right)}{\Gamma\left( \frac{3+j}{4}\right)} M\left(\frac{1+j}{4}, \frac{1}{2}, -X \right)\n\\
&~~~~ -\frac{2X^{\frac{1}{2}}  \cos \left(\frac{1+j}{4}\pi  \right)       }{\Gamma\left( \frac{1+j}{4}\right)}
M\left(\frac{3+j}{4}, \frac{3}{2}, -X \right) 
\Bigg].
\label{eq:Delta-exact}
\end{align}

The non-perturbative character of $\Delta(X)$ can be established directly from the exact expression \eqref{eq:Delta-exact}, without performing a large-$X$ asymptotic expansion. For odd $j$, we distinguish the two cases $j=4m+1$ and $j=4m+3$. For $j=4m+1$, with a non-negative integer $m$, one obtains
\begin{align}
\Delta(X)
&=-2i \sqrt{\pi}\frac{(-1)^m}{\Gamma(m+1)} M\left(m+\frac{1}{2},\frac{1}{2}, -X \right)\\
&=-2i\sqrt{\pi}\frac{(-1)^m}{\Gamma(m+1)} e^{-X}M\left(-m, \frac{1}{2}, X \right).
\end{align}
Here we used the Kummer transformation. 
On the other hand, for $j=4m+3$, one finds
\begin{align}
\Delta(X)
&=4i \sqrt{\pi}
 \frac{ (-1)^{m+1}   X^{\frac{1}{2}}    }{\Gamma\left( m+1\right)}
M\left(m+\frac{3}{2}, \frac{3}{2}, -X \right) \\
&=4i \sqrt{\pi}
 \frac{ (-1)^{m+1}   X^{\frac{1}{2}}    }{\Gamma\left( m+1\right)}
e^{-X}M\left(-m, \frac{3}{2}, X \right).
\end{align}
Since $M(-m,\beta,X)$ is a polynomial in $X$ of degree $m$, for each fixed odd $j$,
$\Delta(X)$ can be written exactly in the form
\begin{align}
\Delta(X)\propto e^{-X}X^{\alpha}\times \mathrm{finite~polynomial~in~}X.
\end{align}
Here $\alpha=0$ for $j=4m+1$ and $\alpha=1/2$ for $j=4m+3$. This gives an exact finite-polynomial representation of the non-perturbative contribution in each fixed odd $j$ local tensor component, rather than an infinite asymptotic expansion.

Thus, the odd part of $T^{\mathrm{cone}}$ provides a purely non-perturbative sector of the initial tensor. From the Borel resummation perspective, the two functions $e^{\pm i\pi\frac{1+j}{4}} U_{\pm}$ in \eqref{eq:Delta-U} correspond to the two lateral resummations of the ordinary perturbative expansion. Therefore $\Delta(X)$ is nothing but the discontinuity of these two Borel resummations.


This finite-polynomial property does not imply that the full tensor trace is a finite polynomial. In the full tensor network, the tensor indices are summed over, and hence $j=a+b+c+d$ is not fixed. Nevertheless, this representation
is conceptually important: the non-perturbative odd sector is already visible
at the level of the local tensor data, before performing the full tensor trace.


\subsubsection{Reality of the tensor trace}
Since the model we are considering is non-Hermitian, the odd $j$ components of the local tensor are purely imaginary. 
However this does not mean that the partition function has a non-vanishing imaginary part. 
In fact, the local tensor satisfies
\begin{align}
\left(T^{\mathrm{cone}}_{abcd}\right)^*
=
(-1)^{a+b+c+d}
T^{\mathrm{cone}}_{abcd}.
\end{align}
For a fixed bond-index configuration in the closed tensor network, the total parity is
\begin{align}
\sum_x (a_x+b_x+c_x+d_x)
=
2\sum_{\ell} n_\ell ,
\end{align}
because each bond index $n_\ell$ is shared by two neighboring tensors. Hence the number of odd $j$ local tensors is always even. Consequently, the imaginary factors from the odd sector appear only in pairs, and each contribution to the tensor trace is real.
The reality of the tensor trace reflects the underlying $\mathcal{PT}$ symmetry.

\subsection{Summary of the cone-contour tensor structure}

Let us summarize the structure revealed by the cone-contour tensor. Although the potential is even in $\phi$, the cone contour is not invariant under the ordinary field-reflection $\mathbb{Z}_2$ transformation $\phi\to-\phi$. For the integration contour over the whole real axis, this $\mathbb{Z}_2$ symmetry would make all odd moments vanish. For the cone contour, however, this cancellation no longer occurs, and the odd-$j$ components of the local tensor become nonzero. In this sense, the cone contour breaks the ordinary $\mathbb{Z}_2$ symmetry of the integration contour while preserving the $\mathcal{PT}$-compatible structure of the formulation.

This is the origin of the parity decomposition found above. The even sector is associated with the symmetric combination of the two lateral resummations and is described by the median resummation of the perturbative sector at the level of each local tensor component. By contrast, the odd sector is associated with the difference between the two lateral resummations. This difference cancels the ordinary perturbative contribution and leaves a purely non-perturbative contribution proportional to $\exp(-A^2/g)$.

A notable feature of the odd sector is that its non-perturbative contribution is obtained directly from the exact expression for the local tensor component, without relying on a large-$X$ asymptotic expansion. For each fixed odd $j$, after extracting the exponential and algebraic factors, the remaining part is exactly a finite polynomial in $X=A^2/g$. Thus, the non-perturbative information associated with the difference between the two lateral resummations is explicitly encoded in each odd-$j$ local tensor component.
The significance of this result is that the non-perturbative odd sector is identified already at the level of the local building blocks of the tensor network.

Finally, although the odd-$j$ components are purely imaginary, they do not make the partition function complex. In a closed tensor network, the number of odd-$j$ tensors is always even because each bond index is shared by two neighboring tensors. Consequently, each bond-index configuration gives a real contribution to the tensor trace, and the cone-contour partition function remains real.

The cone-contour tensor thus separates into a median perturbative even sector and a non-perturbative odd sector at the level of local tensor components. In the next section, we turn to the wedge-contour formulation and establish its relation to the analytic continuation of the Hermitian $\phi^4$ theory.


\section{Wedge contour and the analytic continuation of the Hermitian theory}

We now turn to the wedge-contour formulation of the discretized $\mathcal{PT}$-symmetric $-g\phi^4$ theory. The purpose of this section is different from that of the cone-contour analysis. In the cone-contour formulation, the main object was the parity structure of the local tensor, which separates the median perturbative even sector from the non-perturbative odd sector. In the wedge-contour formulation, by contrast, the main result is an exact analytic relation, formulated in the tensor network representation, between the wedge-contour partition function and the analytic continuation of the Hermitian $\lambda \phi^4$ theory to negative quartic coupling $\lambda<0$.

More precisely, using the tensor network representation, we analytically show that, at finite volume, the wedge-contour partition function can be written as the real part of the analytically continued Hermitian partition function,
\begin{align}
Z_d^{\mathrm{wedge}}(g)
=
\operatorname{Re} Z_d^{H}(\lambda=-g),
\end{align}
in arbitrary spacetime dimension $d$. 
A corresponding relation between the wedge-contour partition function and the real part of the analytically continued Hermitian partition function was established in the one-dimensional setting (i.e., quantum mechanics) in Ref.~\cite{Lawrence:2023woz}. The derivation below formulates this relation in the tensor network representation, first in two dimensions and then in arbitrary spacetime dimensions. 
This relation highlights a different aspect of the $\mathcal{PT}$-symmetric formulation: while the cone contour makes the non-perturbative odd sector visible at the level of local tensor components, the wedge contour gives an exact analytic connection between the $\mathcal{PT}$-symmetric formulation and the lateral analytic continuations of the Hermitian partition function.

\subsection{Wedge-contour partition function}\label{wedgecontour}

We first formulate the wedge-contour partition function in two dimensions for the lattice theory introduced in \eref{lateq}.
The wedge contour is defined by 
\begin{align}
 \begin{split}
\phi_0&=s_0 \left(e^{\frac{\pi i}{4}} \theta(-s_0)+e^{-\frac{\pi i}{4}}   \theta(s_0) \right),\\
\phi_i&=s_i \left(e^{\frac{\pi i}{4}} \theta(-s_0)+e^{-\frac{\pi i}{4}}   \theta(s_0) \right),
 \end{split}
\end{align}
where $s_0\in\mathbb R$ and $s_i\in\mathbb R$ for $i=1,\ldots,V-1$. The difference from the cone contour lies in the arguments of the step functions. For the cone contour, the phase of each $\phi_i$ is determined by the sign of $s_i$ at the same lattice site. In the wedge contour, however, the step functions in all $\phi_i$ depend on the single variable $s_0$. As a result, the partition functions for these distinct contours should be different.

As shown in Ref.~\cite{Lawrence:2023woz}, the wedge-contour partition function is obtained by combining the two branch contributions,
\begin{align}
Z_{d=2}^{\mathrm{wedge}}(g)=
\frac{1}{2}
\left[
Z_{+}^{\mathrm{wedge}}(g)
+
Z_{-}^{\mathrm{wedge}}(g)
\right],
\end{align}
where
\begin{align}
Z^{\mathrm{wedge}}_\pm(g) &=e^{\pm\frac{i\pi  V}{4} }\int_{-\infty}^\infty ds_0 \int_{-\infty}^\infty \prod_{i=1}^{V-1}ds_i\left(e^{-S_\pm[s]}\right).\label{integral2}
\end{align}
In writing \eref{integral2}, we used the global $\mathbb{Z}_2$ symmetry under the simultaneous sign flip $s_i\to -s_i$ at all lattice sites to extend the integration over each branch to the full real domain. The factor $1/2$ in the wedge-contour partition function arises from this extension. The corresponding actions are
\begin{align}
S_\pm[s]
&=
\pm i
\sum_{\langle i,j\rangle}
\frac{(s_i-s_j)^2}{2}
\pm i
\sum_{i\in\Lambda}
\frac{\mu_0^2}{2}s_i^2
+
\sum_{i\in\Lambda}
\frac{g}{4}s_i^4 \label{half2}\\
&=
\sum_{\langle i,j\rangle}
\left[
\pm i
\left(
\frac{1}{2}s_i^2
+
\frac{1}{2}s_j^2
-
s_i s_j
\right)
\pm i
\frac{\mu_0^2}{8}
\left(
s_i^2+s_j^2
\right)
+
\frac{g}{16}
\left(
s_i^4+s_j^4
\right)
\right].
\end{align}
Here, we have rewritten the action as a sum of link contributions. Each local quadratic and quartic term is assigned with a factor of $1/4$ to every link connected to the corresponding site. This form will be used to construct the initial tensor below. As in the previous section, the integral in \eref{integral2} cannot be evaluated directly because of the cross term $s_i s_j$. To cast the partition function into the tensor network formalism, we Taylor-expand this term so that the field integrations factorize site by site
\begin{align}
\exp\left(
\pm i s_i s_j
\right)
=
\sum_{n=0}^{\infty}
\frac{(\pm i)^n}{n!}
s_i^n s_j^n.
\end{align}
This expansion allows us to perform the integrations over $s_i$ and $s_j$ independently. We introduce the one-leg factor as
\begin{align}
f_{\pm, l}(s)
=& \frac{(\pm i)^{\frac{l}{2}} s^l}{\sqrt{l!}} \exp\left[ \mp i\left( \frac{1}{2}+\frac{\mu_0^2}{8}\right)s^2-\frac{g}{16}  s^4 \right].
\end{align}
The initial tensor, which is constructed by the four-legs, is
\begin{align}
T_{\pm, abcd}^{\mathrm{wedge}}
&=\int_{-\infty}^{\infty}ds  \frac{(\pm i)^{\frac{j}{2}} s^{j}}{\sqrt{a! b! c! d!}}
\exp\left[\mp i As^2 -\frac{g}{4}s^4\right]\\
&=
\frac{ (\pm i)^{\frac{j}{2}}   }{\sqrt{a! b! c! d!}}
h^{\mathrm{wedge}}_{\pm}\left(j, A, g\right).\label{wedgetensor}
\end{align}
Here the $h$ function is defined by
\begin{align}
h^{\mathrm{wedge}}_{\pm}\left( j, A, g\right)&\equiv\int_{-\infty}^{\infty}ds s^{j}\exp \left[\mp i As^2-\frac{g}{4}s^4 \right]\label{hfunction0}\\
&=P_j
\left(\frac{1}{g} \right)^{\frac{1+j}{4}}\Gamma\left(\frac{1+j}{2}\right) U \left(\frac{1+j}{4},
 \frac{1}{2}, z_\pm\right)\label{hfunction},
\end{align}
where the parity projector is introduced by
\begin{align}
P_j\equiv\frac{1+(-1)^j}{2}.
\end{align}
Since the integration range is symmetric under $s\to -s$, the defining integral vanishes for odd $j$. Accordingly, the parity projector $P_j$ has been included in \eref{hfunction}.
Using the local tensor in \eref{wedgetensor}, each branch contribution takes the tensor network form
\begin{align}
Z_\pm^{\mathrm{wedge}}(g)
=
e^{
\pm\frac{i\pi V}{4}
}
\mathrm{tTr}
\left[T_\pm^{\mathrm{wedge}}\right].\label{twedge}
\end{align}



\subsection{Relation to the analytic continuation of the Hermitian theory}\label{hermitian}

We now compare the wedge-contour partition function obtained above to the analytic continuation of the Hermitian lattice $\lambda \phi^4$ theory to negative quartic coupling. We begin with the Hermitian lattice theory at positive bare coupling $\lambda>0$, in which all field variables are integrated along the real axis, $\phi_i\in\mathbb{R}$. The positive quartic coupling ensures that the finite-volume lattice integral is convergent. Its partition function in two dimensions is
\begin{align}
Z_{d=2}^{H}(\lambda)
&=
\prod_{i\in\Lambda}\int_{-\infty}^{\infty} d\phi_i
\exp\left[
-S^{H}[\phi;\lambda]
\right],
\end{align}
where
\begin{align}
S^{H}[\phi;\lambda]
&=
\sum_{\langle i,j\rangle}
\frac{(\phi_i-\phi_j)^2}{2}
+
\sum_{i\in\Lambda}
\frac{\mu_0^2}{2}\phi_i^2
+
\sum_{i\in\Lambda}
\frac{\lambda}{4}\phi_i^4\\
&=
\sum_{\langle i,j\rangle}
\left[
\frac{1}{2}\phi_i^2
+
\frac{1}{2}\phi_j^2
-
\phi_i\phi_j
+
\frac{\mu_0^2}{8}
\left(
\phi_i^2+\phi_j^2
\right)
+
\frac{\lambda}{16}
\left(
\phi_i^4+\phi_j^4
\right)
\right].
\end{align}
The second equality expresses the action as a sum of link contributions. Applying the same Taylor expansion of the cross term as in the wedge-contour formulation, we obtain the local tensor
\begin{align}
T_{abcd}^{H}(\lambda)
&=
\frac{1}{\sqrt{a!b!c!d!}}
\int_{-\infty}^{\infty}
d\phi\, \phi^j
\exp\left[
-A
\phi^2-
\frac{\lambda}{4}\phi^4
\right]\\
&=
\frac{h^{H}(j, A,\lambda)}
{\sqrt{a!b!c!d!}},\label{hcon}
\end{align}
where 
\begin{align}
h^{H}(j, A,\lambda)
&=
P_j
\left(
\frac{1}{\lambda}
\right)^{\frac{1+j}{4}}
\Gamma\left(
\frac{1+j}{2}
\right)
U\left(
\frac{1+j}{4},
\frac{1}{2},
\frac{A^2}{\lambda}
\right).
\end{align}
Here the parity projector reflects the vanishing of the real-axis integral for odd $j$.

We now analytically continue the Hermitian local tensor from positive $\lambda$ to negative quartic coupling through the function $h^{H}(j, A,\lambda)$. Since the Tricomi confluent hypergeometric function $U(\alpha,\beta,z)$ appearing in $h^{H}(j, A,\lambda)$ has a branch cut along the negative real axis in the complex $z$-plane, we consider the two lateral continuations
\begin{align}
\lambda
\to
g e^{\pm i(\pi-0)}=
-g\pm i0.
\end{align}
Then,
\begin{align}
h^{H}(j, A,\lambda)
\to
h^{H}_{\pm}
\left(
j, A,
g e^{\pm i(\pi-0)}
\right),
\end{align}
where 
\begin{align}
h^{H}_{\pm}(j, A, g e^{\pm i (\pi-0)})
 &=P_j e^{\mp \frac{i\pi}{4}}(\mp i)^{\frac{j}{2}} \left(\frac{1}{g} \right)^{\frac{1+j}{4}}\Gamma\left(\frac{1+j}{2}\right)U \left(\frac{1+j}{4},
 \frac{1}{2}, z_\mp \right).
\end{align}
Substituting $h^{H}_{\pm}$ into \eref{hcon}, we obtain the analytically continued initial tensor
\begin{align}
T^{H}_{\pm,abcd}(g)
&\equiv
T^{H}_{abcd}
\left(
g e^{\pm i(\pi-0)}
\right)\\
&=
P_j
e^{\mp \frac{i\pi}{4}}
\frac{(\mp i)^{\frac{j}{2}}}{\sqrt{a!b!c!d!}}
\left(
\frac{1}{g}
\right)^{\frac{1+j}{4}}
\Gamma\left(
\frac{1+j}{2}
\right)
U\left(
\frac{1+j}{4},
\frac{1}{2},
z_\mp
\right).\label{84}
\end{align}

We also note a relation to the cone-contour construction. Comparing \eref{39} and \eref{84}, we obtain
\begin{align}
T^{H}_{\pm,abcd}(g)
=
P_j \tilde{T}_{\pm,abcd}.
\end{align}
Hence, as seen from \eref{40}, the cone-contour initial tensor is a parity-dependent linear combination of the expressions corresponding to the two laterally continued Hermitian initial tensors, with the projection factor $P_j$ omitted.

The corresponding analytically continued partition functions are
\begin{align}
Z^{H }_\pm(g)&=\mathrm{tTr}\left[T^{H}_\pm (g)\right]\\
&=e^{\mp\frac{ i\pi V}{4}}\mathrm{tTr}\left[ 
P_j
\frac{(\mp i)^{\frac{j}{2}}}{\sqrt{a!b!c!d!}}
\left(
\frac{1}{g}
\right)^{\frac{1+j}{4}}
\Gamma\left(
\frac{1+j}{2}
\right)
U\left(
\frac{1+j}{4},
\frac{1}{2},
z_\mp
\right)
 \right].
\end{align}

For comparison, we rewrite the two branch contributions to the wedge-contour partition function \eref{twedge} as
\begin{align}
Z^{\mathrm{wedge}}_\pm (g)&= e^{\pm\frac{ i \pi  V}{4}}\mathrm{tTr} 
\left[P_j\frac{ (\pm i)^{\frac{j}{2}}   }{\sqrt{a! b! c! d!}}\left(\frac{1}{g} \right)^{\frac{1+j}{4}}\Gamma\left(\frac{1+j}{2}\right) U \left(\frac{1+j}{4},
 \frac{1}{2}, z_\pm\right)\right]
\end{align}
Comparing the two tensor network expressions, we obtain the important relation
\begin{align}
Z^{\mathrm{wedge}}_\pm(g)
=Z^H_\mp(g).
\end{align}
For real $g$ and $\mu_0^2$, the two lateral analytic continuations from the positive-$\lambda$ Hermitian theory are complex conjugates,
\begin{align}
Z^H_-(g)
=
\left[
Z^H_+(g)
\right]^*.
\end{align}
Therefore, the average of the two lateral continuations is equal to the real part of either one:
\begin{align}
Z_{d=2}^{\mathrm{wedge}}(g)&=
\frac{1}{2}
\left[
Z_+^{\mathrm{wedge}}(g)+
Z_-^{\mathrm{wedge}}(g)
\right]\\
&=
\frac{1}{2}
\left[
Z_-^{H}(g)
+
Z_+^{H}(g)
\right]\\
&=
\operatorname{Re}
Z^{H}_{+}(g),
\end{align}
where we denote this real quantity by
\begin{align}
\operatorname{Re}
Z_{d=2}^{H}(\lambda=-g)
\equiv
\operatorname{Re}
Z_+^{H}(g)
=
\operatorname{Re}
Z_-^{H}(g).
\end{align}
Hence,
\begin{align}
Z_{d=2}^{\mathrm{wedge}}(g)
=\operatorname{Re}
Z_{d=2}^{H}(\lambda=-g).
\end{align}

\subsection{Extension to arbitrary dimensions}
Building on the two-dimensional analysis in Sec.~\ref{wedgecontour} and Sec.~\ref{hermitian}, we show that the relation between the wedge-contour partition function and the real part of the analytically continued Hermitian partition function persists in arbitrary spacetime dimensions. The argument concerns the analytic structure of the local tensors; the numerical contraction of higher-dimensional tensor networks is beyond the scope of this work.


On a $d$-dimensional hypercubic lattice, the local tensor carries $2d$ indices. For later use, we introduce the notation:
\begin{align}
J&\equiv \sum_{i=1}^{2d} n_i, \quad A_d\equiv d+\frac{\mu_0^2}{2}>0, \quad z_{d,\pm}\equiv\frac{A_d^2}{g}e^{\pm i(\pi-0)}.
\end{align}
Using the same wedge-contour prescription as in the two-dimensional case, the local tensor is
\begin{align}
T^{\mathrm{wedge}}_{d, \pm, n_1 n_2\cdots n_{2d-1}n_{2d}}
&=\frac{  (\pm i)^{\frac{J}{2}}} {\prod_{i=1}^{2d}\sqrt{n_i !}} h^{\mathrm{wedge}}_{d, \pm}\left(J, A_d, g\right),
\end{align}
where
\begin{align}
h^{\mathrm{wedge}}_{d, \pm}\left(J, A_d, g\right)&\equiv 
\int_{-\infty}^{\infty}ds s^{J}\exp \left[\mp iA_d s^2-\frac{g}{4}s^4 \right]\\
&=P_J \left(\frac{1}{g} \right)^{\frac{1+J}{4}}
\Gamma\left(\frac{1+J}{2} \right)U\left(\frac{1+J}{4}, \frac{1}{2}, z_{d,\pm}\right),
\end{align}
where
\begin{align}
P_J\equiv\frac{1+(-1)^J}{2}
\end{align}
is the parity projector.
Writing the volume of the periodic lattice as $V=N^d$, the two branch contributions to the wedge-contour partition function are
\begin{align}
Z^{\mathrm{wedge}}_{d,\pm}(g)
&=
e^{\pm\frac{i\pi V}{4}}
\mathrm{tTr}
\left[
T^{\mathrm{wedge}}_{d,\pm}
\right]\\
&= 
 e^{\pm\frac{i\pi V}{4}}   
\mathrm{tTr} \left[\frac{  (\pm i)^{\frac{J}{2}}} {\prod_{i=1}^{2d}\sqrt{n_i !}} P_J \left(\frac{1}{g} \right)^{\frac{1+J}{4}}
\Gamma\left(\frac{1+J}{2} \right)U\left(\frac{1+J}{4}, \frac{1}{2}, z_{d,\pm}\right)\right].\label{dwed2}
\end{align}

On the other hand, the initial tensor of the Hermitian theory at positive quartic coupling is
\begin{align}
T^H_{d,n_1\cdots n_{2d}}(\lambda)
&=
\frac{
h_d^H(J,A_d,\lambda)
}
{\prod_{i=1}^{2d}\sqrt{n_i!}},
\end{align}
where
\begin{align}
h^H_d\left( J, A_d, \lambda\right)&\equiv\int_{-\infty}^{\infty}d\phi \phi^{J}\exp \left[-A_d\phi^2-\frac{\lambda}{4}\phi^4 \right]\\
&=P_J\left(\frac{1}{\lambda}\right)^{\frac{1+J}{4}}\Gamma\left(\frac{1+J}{2} \right)U\left(\frac{1+J}{4}, \frac{1}{2}, 
\frac{A_d^2}{\lambda}\right).
\end{align}
When we analytically continue the coupling as $\lambda\rightarrow g e^{\pm i(\pi-0)}=-g\pm i 0$, the two lateral continuations of the local function are 
\begin{align}
h^H_{d,\pm}(J,A_d,g)
&\equiv
h_d^H
\left(
J,A_d,
g e^{\pm i(\pi-0)}
\right)
\\
&=
P_J
e^{\mp\frac{i\pi}{4}}
(\mp i)^{\frac{J}{2}}
\left(
\frac{1}{g}
\right)^{\frac{1+J}{4}}
\Gamma\left(
\frac{1+J}{2}
\right)
U\left(
\frac{1+J}{4},
\frac{1}{2},
z_{d,\mp}
\right),
\end{align}
and the analytically continued initial tensor is
\begin{align}
T^H_{d,\pm, n_1\cdots n_{2d}}(g)&\equiv T^H_{d, n_1\cdots n_{2d}}\left(g e^{\pm i(\pi-0)}\right)\\
&=P_J e^{\mp \frac{i\pi}{4}}
\frac{(\mp i)^{\frac{J}{2}}}{\prod_{i=1}^{2d}\sqrt{n_i !}} 
\left(\frac{1}{g} \right)^{\frac{1+J}{4}} \Gamma\left(\frac{1+J}{2}\right)U \left(\frac{1+J}{4},
 \frac{1}{2}, z_{d,\mp}\right).
\end{align}
The corresponding analytically continued partition functions are
\begin{align}
Z^H_{d,\pm}(g)
&=
\mathrm{tTr}
\left[
T^H_{d,\pm}(g)
\right]\\
&=
e^{\mp\frac{i\pi V}{4}}
\mathrm{tTr}
\left[
P_J
\frac{(\mp i)^{\frac{J}{2}}}
{\prod_{i=1}^{2d}\sqrt{n_i!}}
\left(
\frac{1}{g}
\right)^{\frac{1+J}{4}}
\Gamma\left(
\frac{1+J}{2}
\right)
U\left(
\frac{1+J}{4},
\frac{1}{2},
z_{d,\mp}
\right)
\right].
\label{dh2}
\end{align}
Comparing \eref{dh2} with \eref{dwed2}, we obtain
\begin{align}
Z^H_{d,\pm}(g)
=
Z^{\mathrm{wedge}}_{d,\mp}(g).
\end{align}


For real $g$ and $\mu_0^2$, the two lateral continuations are complex conjugates of each other:
\begin{align}
Z^H_{d,-}(g)
=\left[Z^H_{d,+}(g)\right]^*.
\end{align}
Therefore,
\begin{align}
Z_d^{\mathrm{wedge}}(g)
&\equiv
\frac{1}{2}
\left[
Z^{\mathrm{wedge}}_{d,+}(g)
+
Z^{\mathrm{wedge}}_{d,-}(g)
\right]\\
&=
\frac{1}{2}
\left[
Z^H_{d,-}(g)
+
Z^H_{d,+}(g)
\right]\\
&=
\operatorname{Re}
Z^H_{d,+}(g).
\end{align}
We define the real part of the Hermitian partition function at negative quartic coupling by
\begin{align}
\operatorname{Re}
Z_d^H(\lambda=-g)
&\equiv
\operatorname{Re}
Z^H_{d,+}(g)
=
\operatorname{Re}
Z^H_{d,-}(g).
\end{align}
We have thus established, for a finite periodic lattice in arbitrary spacetime dimensions, the desired relation
\begin{align}
Z_d^{\mathrm{wedge}}(g)
&=
\operatorname{Re}
Z_d^H(\lambda=-g).
\label{desire}
\end{align}



\section{Summary and discussion}

In this paper, we developed a tensor network formulation of scalar field
theories whose path integrals are defined on complex integration contours.
After lattice discretization, the dependence on the integration contour is
encoded in a one-variable integral defining the local initial tensor.

Applying this framework to the lattice $-g\phi^4$ theory, we obtained two
main results associated with two distinct contour prescriptions. First, for
the two-dimensional $\mathcal{PT}$-symmetric theory defined on the cone
contour, we derived an explicit analytic expression for the initial tensor.
The tensor separates into even and odd sectors according to the parity of
the sum of its indices. The even-sector components are real and admit
perturbative asymptotic expansions. From the viewpoint of Borel
resummation, their exact analytic expressions are given by the median
resummations of these expansions. By contrast, the odd-sector components
are purely imaginary and contain exponentially suppressed non-perturbative
contributions. For each fixed odd $j$, after factoring out the exponential
and algebraic factors, the remaining dependence is exactly
a finite polynomial in $X=A^2/g$. Although the individual odd-sector
components are purely imaginary, every closed-network contraction contains
an even number of them, ensuring that the cone-contour partition function
remains real. Thus, the non-perturbative structure is already visible in the
local building blocks of the tensor network before the full tensor trace is
performed. This result shows that the tensor network method is useful not only as a numerical tool, but also as an analytic framework for identifying non-perturbative structures.

Second, for an alternative wedge-contour prescription, we established an
exact finite-volume relation on a periodic hypercubic lattice in arbitrary
dimension $d$. The wedge-contour partition function is equal to the average
of the two lateral analytic continuations of the Hermitian lattice
$\lambda\phi^4$ partition function to negative quartic coupling, or
equivalently to the real part of either lateral continuation. We first derived
this relation in two dimensions using the tensor network representation and
then showed that it persists in arbitrary dimensions at the level of the
corresponding local tensors. This wedge-contour prescription should be
distinguished from the cone-contour definition of the
$\mathcal{PT}$-symmetric theory.

A practical outcome of the present work is the explicit construction of the
cone-contour initial tensor, which provides concrete input for numerical
tensor network calculations. In two dimensions, this tensor can be used
with tensor network coarse-graining methods to investigate the phase
structure and critical behavior of the cone-contour theory. A key problem
for future numerical work is to determine how the non-perturbative contributions encoded in
the local tensor components are reflected in global quantities such as the
partition function, free energy, and critical observables. A further question
is whether and how the cone-contour partition function is related to the
analytic continuation of the Hermitian lattice $\lambda\phi^4$ partition
function. Finally, the present framework can be extended to other
$\mathcal{PT}$-symmetric field theories with different interactions in order
to investigate their non-perturbative properties using tensor network
methods.

\section*{Acknowledgement}
We would like to thank Daisuke Kadoh and Shinji Takeda for useful discussions. This work was supported by JSPS KAKENHI Grants Nos. 	23K25790, 26K07110 (Y.H.) and 25K23384 (K.K.).


\appendix


\section{Tensor trace}\label{tensortrace}

For a two-dimensional tensor network, the tensor trace
$\mathrm{tTr}$ denotes the contraction of all nearest-neighbor
tensor indices.
We assign the four indices of the local tensor as
\begin{align}
T_{abcd}=T_{\mathrm{right},\mathrm{up},\mathrm{left},\mathrm{down}} .
\end{align}
Then the partition function is written as
\begin{align}
Z_{\mathcal C}
&=
\mathrm{tTr}[T]
\nonumber\\
&=
\sum_{\{a_x,b_x,c_x,d_x\}}
\prod_x
T_{a_x b_x c_x d_x}
\prod_x
\delta_{a_x,c_{x+\hat1}}
\delta_{b_x,d_{x+\hat2}} .
\end{align}
Equivalently, introducing a link index $n_{x,\mu}$ on each lattice link
$(x,x+\hat\mu)$, one obtains
\begin{align}
Z_{\mathcal C}
=
\sum_{\{n_{x,\mu}\}}
\prod_x
T_{
n_{x,\hat1}\,
n_{x,\hat2}\,
n_{x-\hat1,\hat1}\,
n_{x-\hat2,\hat2}
}.
\end{align}

The same notation extends straightforwardly to arbitrary dimensions.
In $d$ dimensions, we order the $2d$ indices of the local tensor as
\begin{align}
T_{n_1,\ldots,n_{2d}}
=
T_{+\hat1,\ldots,+\hat d,-\hat1,\ldots,-\hat d}.
\end{align}
The partition function is then written as
\begin{align}
Z_{\mathcal C}
=
\sum_{\{n_{x,\mu}\}}
\prod_x
T_{
n_{x,\hat1}
\ldots
n_{x,\hat d}
n_{x-\hat1,\hat1}
\ldots
n_{x-\hat d,\hat d}
},
\end{align}
where $\mu=1,\ldots,d$ labels the lattice directions.
The sum over the indices on all lattice links is
explicitly defined as
\begin{align}
\sum_{\{n_{x,\mu}\}}
\equiv
\prod_{x\in\Lambda}
\prod_{\mu=1}^{d}
\left(
\sum_{n_{x,\mu}=0}^{\infty}
\right).
\end{align}


\section{Confluent hypergeometric functions}

We use the notation of the DLMF for confluent hypergeometric functions
\cite{NIST:DLMF0}. The Kummer confluent hypergeometric function is defined by
\begin{align}
M(a,b,z)
= {}_1F_1(a;b;z)=
\sum_{k=0}^{\infty}
\frac{(a)_k}{(b)_k}
\frac{z^k}{k!}.
\end{align}
This defining series is DLMF Eq.~13.2.2. Here, $(a)_k$ denotes the Pochhammer symbol,
\begin{align}
(a)_0=1,
\qquad
(a)_k=a(a+1)\cdots(a+k-1)
\quad (k\geq1).
\end{align}
In particular, when $a=-m$ with $m$ a non-negative integer, one has
\begin{align}
(-m)_k=0
\qquad (k>m),
\end{align}
so that $M(-m,b,z)$ is a polynomial in $z$ of degree $m$.



We also use the connection formula between the Tricomi confluent
hypergeometric function $U(a,b,z)$ and the Kummer function $M(a,b,z)$.
For $b\notin \mathbb{Z}$, it is given by
\begin{align}
U(a,b,z)
=
\frac{\Gamma(1-b)}{\Gamma(a-b+1)}
M(a,b,z)
+
\frac{\Gamma(b-1)}{\Gamma(a)}
z^{1-b}
M(a-b+1,2-b,z).
\end{align}
This corresponds to DLMF Eq.~13.2.41.
We apply this formula to
\begin{align}
U_\pm
\equiv
U\left(
\frac{1+j}{4},
\frac{1}{2},
z_\pm
\right),
\end{align}
where
\begin{align}
z_\pm \equiv X e^{\pm i (\pi-0)} = -X\pm i0, \qquad X \equiv \frac{A^2}{g}, \qquad A \equiv 2+\frac{\mu_0^2}{2}.
\end{align}
Using $z_\pm^{1/2}=\pm iX^{1/2}$ on the principal branch and the fact
that $M(a,b,z)$ is entire in $z$, we obtain
\begin{align}
U_\pm
&=
U\left(\frac{1+j}{4},\frac{1}{2}, X e^{\pm i(\pi-0)}  \right)
\nonumber\\
&=
\frac{\sqrt{\pi}}{\Gamma\left(\frac{3+j}{4}\right)}
M\left(\frac{1+j}{4},\frac{1}{2}, -X  \right)
\mp iX^{\frac{1}{2}}
\frac{2\sqrt{\pi}}{\Gamma\left(\frac{1+j}{4}\right)}
M\left(\frac{3+j}{4}, \frac{3}{2}, -X\right).
\end{align}

For the large-$X$ analysis of the even sector in the main text, we need the
asymptotic forms of both $U(a,b,z)$ and $M(a,b,z)$ near the negative real axis. With
$z_\pm=-X\pm i0$, the large-argument expansion of the Tricomi and Kummer functions
can be written as
\begin{align}
U(a,b,z_\pm)
\sim
(-X)^{-a}
\sum_{s=0}^{\infty}
\frac{(a)_s(a-b+1)_s}{s!}
X^{-s},
\end{align}
which corresponds to DLMF Eq.~13.7.3, and
\begin{align}
M(a,b,z_\pm)
\sim&
\frac{\Gamma(b)e^{z_\pm}z_\pm^{a-b}}{\Gamma(a)}
\sum_{s=0}^{\infty}
\frac{(1-a)_s(b-a)_s}{s!}
z_\pm^{-s}
\nonumber\\
&+
\frac{e^{\pm i\pi a}\Gamma(b)z_\pm^{-a}}{\Gamma(b-a)}
\sum_{s=0}^{\infty}
\frac{(a)_s(a-b+1)_s}{s!}
(-z_\pm)^{-s},
\end{align}
which corresponds to DLMF Eq.~13.7.2.

We also use the following exact identity in the analysis of the odd sector.
Using Kummer's transformation, DLMF Eq.~13.2.39, we have for $X>0$
\begin{align}
M(a,b,-X)
=
e^{-X}M(b-a,b,X).
\end{align}

\section{Numerical tests in quantum mechanics}

In this appendix, we present numerical tests of our tensor network formulation in quantum mechanics, corresponding to $d=1$.
Since no renormalization is required in quantum mechanics, the numerical analysis is considerably simpler than in $d\geq 2$.
The $d=1$ theory with a pure, or massless, quartic potential was analyzed in detail in Ref.~\cite{Lawrence:2023woz}.
Here, we apply our tensor network formulation to the standard quartic oscillator with the Euclidean action
\begin{align}
S[x]
=
\int\! d\tau\,
\biggl[
\frac{1}{2}(\partial_\tau x)^2
+\frac{1}{2}x^2
+\frac{\lambda}{4}x^4
\biggr],
\end{align}
where we have set $\hbar=m_0=1$ and renamed the field variable $\phi$ as $x$.
For $\lambda>0$, the corresponding quantum theory is stable and Hermitian.
For $\lambda=-g<0$, by contrast, the potential is unbounded from below along the real $x$ axis.

The corresponding Hamiltonian is
\begin{align}
H
=
\frac{1}{2}p^2+\frac{1}{2}x^2+\frac{\lambda}{4}x^4.
\label{eq:H-quartic}
\end{align}
For $\lambda=-g<0$, one may consider two distinct eigenvalue problems.

The first is the resonance problem.
Along the real axis, $x\in\mathbb{R}$, the potential has a metastable well around $x=0$.
It is therefore natural to impose the Gamow--Siegert boundary conditions, which require purely outgoing behavior as $x\to\pm\infty$.
These boundary conditions give rise to a discrete set of complex eigenvalues, known as resonance energies, which we denote by $E_n^{\mathrm{res}}(g)$.
It is well known that these resonance energies can be obtained by analytically continuing the bound state energies $E_n^H(\lambda)$ of the stable Hermitian oscillator from $\lambda>0$ to the upper edge of the negative coupling axis~\cite{Bender:1969si,Bender:1973rz,Jentschura:2010zza}:
\begin{align}
E_n^{\mathrm{res}}(g)
=
E_n^{H}(\lambda=-g+i0),
\qquad
\operatorname{Im}E_n^{\mathrm{res}}(g)<0.
\end{align}
Consequently, the corresponding partition function is given by
\begin{align}
Z_{d=1}^{\mathrm{res}}(g)
=
Z_{d=1}^{H}(\lambda=-g+i0),
\end{align}
where
\begin{align}
Z_{d=1}^{\mathrm{res}}(g)
=
\sum_{n=0}^{\infty}
e^{-\beta E_n^{\mathrm{res}}(g)},
\qquad
Z_{d=1}^{H}(\lambda)
=
\sum_{n=0}^{\infty}
e^{-\beta E_n^{H}(\lambda)}.
\label{eq:Z-res}
\end{align}

The second is the $\mathcal{PT}$-symmetric eigenvalue problem.
In a $\mathcal{PT}$-symmetric quantum-mechanical model, the coordinate $x$ is complexified, and the wave function is required to satisfy an $L^2$ boundary condition along an appropriate contour in the complex $x$ plane.
The two asymptotic ends of the contour lie within a pair of $\mathcal{PT}$-symmetric Stokes wedges in which the wave function decays exponentially.
Here, we adopt the contour used in Ref.~\cite{Bender:1998ke}.
These boundary conditions select a discrete set of real eigenvalues, which we denote by $E_n^{\mathcal{PT}}(g)$.
The corresponding $\mathcal{PT}$-symmetric partition function is
\begin{align}
Z_{d=1}^{\mathcal{PT}}(g)
=
\sum_{n=0}^{\infty}
e^{-\beta E_n^{\mathcal{PT}}(g)}.
\label{eq:Z-PT}
\end{align}
Thus, although the resonance and $\mathcal{PT}$-symmetric problems are defined by the same formal Hamiltonian \eqref{eq:H-quartic} with $\lambda=-g<0$, they correspond to different boundary conditions and therefore have, in general, different spectra and partition functions.

For a lattice with $N$ sites, the tensor network partition functions are given by
\begin{align}
Z_{d=1,N}^{\mathrm{cone}}(g)
&=
\operatorname{Tr}[
(T^{\mathrm{cone}})^N
],
\\
Z_{d=1,N}^{\mathrm{wedge}}(g)
&=
\frac{
e^{\frac{i\pi N}{4}}
\operatorname{Tr}[
(T_+^{\mathrm{wedge}})^N
]
+
e^{-\frac{i\pi N}{4}}
\operatorname{Tr}[
(T_-^{\mathrm{wedge}})^N
]
}{2},
\end{align}
where
\begin{align}
T_{ab}^{\mathrm{cone}}
&=
\frac{1}{2\sqrt{2\pi\epsilon^{1+J}a!b!}}
\biggl(\frac{1}{\epsilon g}\biggr)^{\frac{1+J}{4}}
\Gamma\biggl(\frac{1+J}{2}\biggr)
\left[
e^{-i\pi\frac{1+J}{4}}U_-
+
(-1)^J e^{i\pi\frac{1+J}{4}}U_+
\right],
\\
T_{\pm,ab}^{\mathrm{wedge}}
&=
\frac{1+(-1)^J}{2}
\frac{e^{\pm i\pi\frac{J}{4}}}
{\sqrt{2\pi\epsilon^{1+J}a!b!}}
\biggl(\frac{1}{\epsilon g}\biggr)^{\frac{1+J}{4}}
\Gamma\biggl(\frac{1+J}{2}\biggr)
U_{\pm},
\end{align}
with $J=a+b$ and
\begin{align}
U_\pm
&=
U\left(
\frac{1+J}{4},
\frac{1}{2},
\frac{A_1^2}{\epsilon g}e^{\pm i(\pi-0)}
\right),
&
A_1
&=
\frac{1}{\epsilon}+\frac{\epsilon}{2}.
\end{align}
Here, we have explicitly displayed the dependence on the lattice spacing $\epsilon=\beta/N$.

Our tensor network formulation predicts that the following relations hold in the continuum limit:
\begin{align}
\lim_{N\to\infty}
Z_{d=1,N}^{\mathrm{cone}}(g)
&=
Z_{d=1}^{\mathcal{PT}}(g),
\\
\lim_{N\to\infty}
Z_{d=1,N}^{\mathrm{wedge}}(g)
&=
\operatorname{Re}
Z_{d=1}^{H}(\lambda=-g)
=
\operatorname{Re}
Z_{d=1}^{\mathrm{res}}(g).
\end{align}
To test how accurately these continuum relations are reproduced at finite lattice spacing, we evaluate the partition functions for $N=40$ and $\beta=5$, corresponding to $\epsilon=\beta/N=1/8$. The truncation sizes required for the transfer matrices $T_{ab}^{\mathrm{cone}}$ and $T_{\pm,ab}^{\mathrm{wedge}}$ depend on the coupling constant $g$ and the lattice spacing $\epsilon$. For each value of $g$, we choose a sufficiently large truncation size such that the resulting partition function has converged to a stable value.

To evaluate the corresponding continuum partition functions, we first compute the relevant energy eigenvalues. The resonance energies are obtained using the complex scaling method \cite{Aguilar:1971ve, Balslev:1971vb}. For the $\mathcal{PT}$-symmetric quartic oscillator, we instead use the equivalent dual Hermitian model with Hamiltonian \cite{Jones:2006qs, Bender:2006wt}
\begin{align}
H_{\mathrm{dual}}
=
\frac{1}{2}p^2
-\sqrt{\frac{g}{2}}x
+g\biggl(x^2-\frac{1}{2g}\biggr)^2,
\end{align}
whose eigenvalues can be efficiently computed using a spectral method.

\begin{figure}[tb]
  \begin{minipage}[b]{0.45\linewidth}
    \centering
    \includegraphics[width=0.95\linewidth]{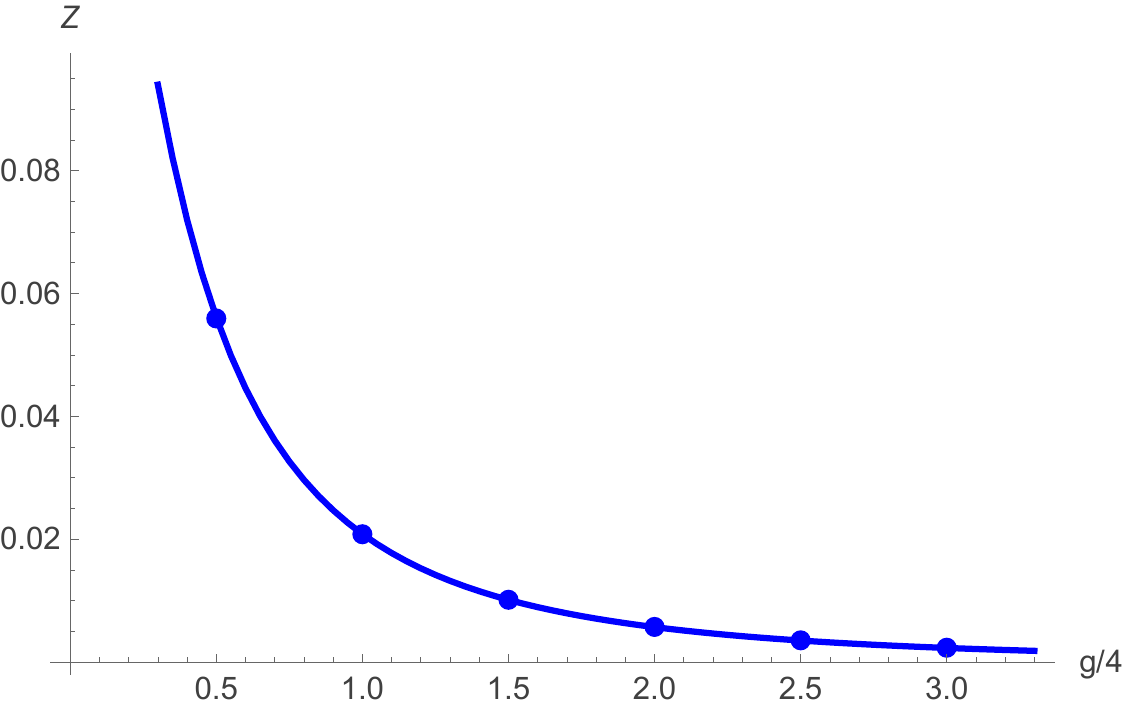}
  \end{minipage}
  \begin{minipage}[b]{0.45\linewidth}
    \centering
    \includegraphics[width=0.95\linewidth]{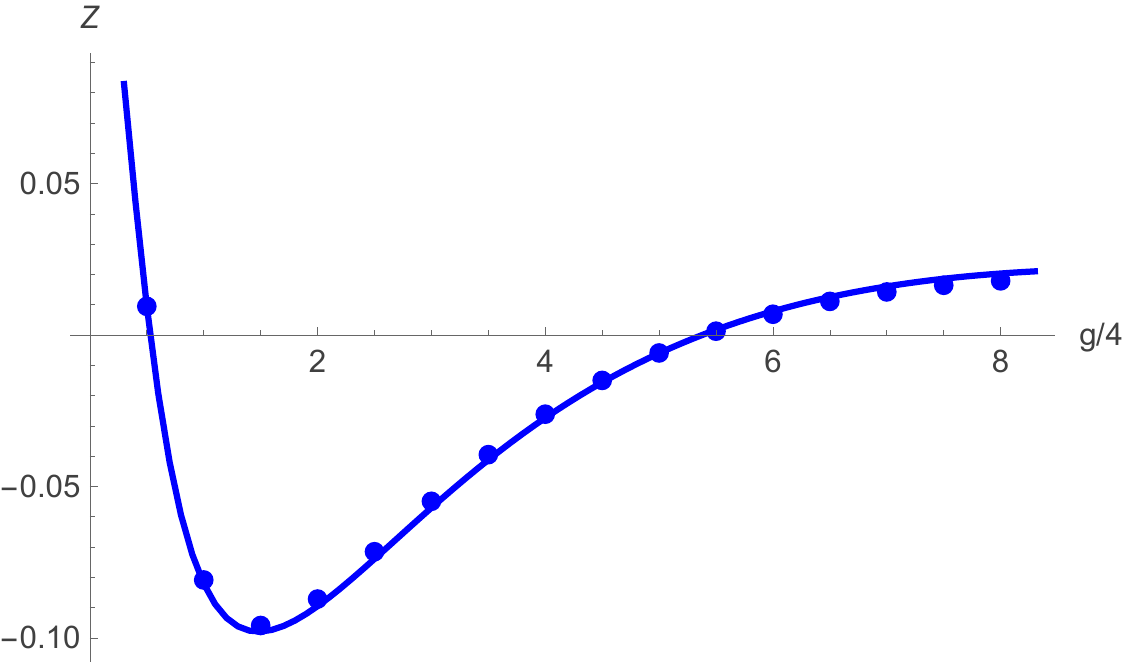}
  \end{minipage}
  \caption{
  Left: The solid curve represents the partition function
  $Z_{d=1}^{\mathcal{PT}}(g)$ of the $\mathcal{PT}$-symmetric quartic oscillator, computed from Eq.~\eqref{eq:Z-PT}, while the dots show the lattice partition function
  $Z_{d=1,N}^{\mathrm{cone}}(g)$ computed using the transfer matrix
  $T_{ab}^{\mathrm{cone}}$.
  Right: The solid curve represents the real part of the resonance partition function
  $Z_{d=1}^{\mathrm{res}}(g)$, computed from Eq.~\eqref{eq:Z-res}, while the dots show the lattice partition function
  $Z_{d=1,N}^{\mathrm{wedge}}(g)$ computed using the transfer matrices
  $T_{\pm,ab}^{\mathrm{wedge}}$.
  The numerical results were obtained for $N=40$ and $\beta=5$.
  }
  \label{fig:Z-comparison}
\end{figure}

The results are shown in Fig.~\ref{fig:Z-comparison}. Even for the finite lattice size $N=40$, the tensor network results are already in good agreement with the corresponding continuum values.

\bibliographystyle{utphys}

\providecommand{\href}[2]{#2}\begingroup\raggedright\endgroup


\begin{thebibliography}{10}

\bibitem{Bender:1998ke}
C.~M. Bender and S.~Boettcher, ``{Real Spectra in Non-Hermitian Hamiltonians
  Having PT Symmetry},''
  \href{http://dx.doi.org/10.1103/PhysRevLett.80.5243}{{\em Phys. Rev. Lett.}
  {\bfseries 80} (1998) 5243--5246},
  \href{http://arxiv.org/abs/physics/9712001}{{\ttfamily
  arXiv:physics/9712001}}.

\bibitem{Bender:2007nj}
C.~M. Bender, ``{Making Sense of Non-Hermitian Hamiltonians},''
  \href{http://dx.doi.org/10.1088/0034-4885/70/6/R03}{{\em Rept. Prog. Phys.}
  {\bfseries 70} (2007) 947},
  \href{http://arxiv.org/abs/hep-th/0703096}{{\ttfamily arXiv:hep-th/0703096}}.

\bibitem{Fisher:1978pf}
M.~E. Fisher, ``{Yang-Lee Edge Singularity and $\phi^3$ Field Theory},''
  \href{http://dx.doi.org/10.1103/PhysRevLett.40.1610}{{\em Phys. Rev. Lett.}
  {\bfseries 40} (1978) 1610--1613}.

\bibitem{Bender:2012ea}
C.~M. Bender, V.~Branchina, and E.~Messina, ``{Ordinary versus PT-symmetric
  $\phi^3$ quantum field theory},''
  \href{http://dx.doi.org/10.1103/PhysRevD.85.085001}{{\em Phys. Rev. D}
  {\bfseries 85} (2012) 085001},
  \href{http://arxiv.org/abs/1201.1244}{{\ttfamily arXiv:1201.1244 [hep-th]}}.

\bibitem{Bender:2013qp}
C.~M. Bender, V.~Branchina, and E.~Messina, ``{Critical behavior of the
  PT-symmetric $i\phi^3$ quantum field theory},''
  \href{http://dx.doi.org/10.1103/PhysRevD.87.085029}{{\em Phys. Rev. D}
  {\bfseries 87} no.~8, (2013) 085029},
  \href{http://arxiv.org/abs/1301.6207}{{\ttfamily arXiv:1301.6207 [hep-th]}}.

\bibitem{Lencses:2024wib}
M.~Lencs{\'e}s, A.~Miscioscia, G.~Mussardo, and G.~Tak{\'a}cs,
  ``{Ginzburg-Landau description for multicritical Yang-Lee models},''
  \href{http://dx.doi.org/10.1007/JHEP08(2024)224}{{\em JHEP} {\bfseries 08}
  (2024) 224}, \href{http://arxiv.org/abs/2404.06100}{{\ttfamily
  arXiv:2404.06100 [cond-mat.stat-mech]}}.

\bibitem{Katsevich:2025ojk}
A.~Katsevich, I.~R. Klebanov, Z.~Sun, and G.~Tarnopolsky, ``{Towards a Quintic
  Ginzburg-Landau Description of the (2,7) Minimal Model},''
  \href{http://dx.doi.org/10.1103/9zz4-pvp7}{{\em Phys. Rev. Lett.} {\bfseries
  136} no.~11, (2026) 111602},
  \href{http://arxiv.org/abs/2510.19085}{{\ttfamily arXiv:2510.19085
  [hep-th]}}.

\bibitem{Bender:1999ek}
C.~M. Bender, K.~A. Milton, and V.~M. Savage, ``{Solution of Schwinger-Dyson
  equations for PT symmetric quantum field theory},''
  \href{http://dx.doi.org/10.1103/PhysRevD.62.085001}{{\em Phys. Rev. D}
  {\bfseries 62} (2000) 085001},
  \href{http://arxiv.org/abs/hep-th/9907045}{{\ttfamily arXiv:hep-th/9907045}}.

\bibitem{Bender:2001qx}
C.~M. Bender, S.~Boettcher, H.~F. Jones, P.~N. Meisinger, and M.~Simsek,
  ``{Bound States of Non-Hermitian Quantum Field Theories},''
  \href{http://dx.doi.org/10.1016/S0375-9601(01)00745-9}{{\em Phys. Lett. A}
  {\bfseries 291} (2001) 197--202},
  \href{http://arxiv.org/abs/hep-th/0108057}{{\ttfamily arXiv:hep-th/0108057}}.

\bibitem{Ai:2022csx}
W.~Ai, C.~M. Bender, and S.~Sarkar, ``{PT-symmetric $-g\varphi^4$ theory},''
  \href{http://dx.doi.org/10.1103/PhysRevD.106.125016}{{\em Phys. Rev. D}
  {\bfseries 106} no.~12, (2022) 125016},
  \href{http://arxiv.org/abs/2209.07897}{{\ttfamily arXiv:2209.07897
  [hep-th]}}.

\bibitem{Lawrence:2023woz}
S.~Lawrence, R.~Weller, C.~Peterson, and P.~Romatschke, ``{Instantons, analytic
  continuation, and PT-symmetric field theory},''
  \href{http://dx.doi.org/10.1103/PhysRevD.108.085013}{{\em Phys. Rev. D}
  {\bfseries 108} no.~8, (2023) 085013},
  \href{http://arxiv.org/abs/2303.01470}{{\ttfamily arXiv:2303.01470
  [hep-th]}}.

\bibitem{Kadoh:2019ube}
D.~Kadoh, Y.~Kuramashi, Y.~Nakamura, R.~Sakai, S.~Takeda, and Y.~Yoshimura,
  ``{Investigation of complex $\phi^{4}$ theory at finite density in two
  dimensions using TRG},''
  \href{http://dx.doi.org/10.1007/JHEP02(2020)161}{{\em JHEP} {\bfseries 02}
  (2020) 161}, \href{http://arxiv.org/abs/1912.13092}{{\ttfamily
  arXiv:1912.13092 [hep-lat]}}.

\bibitem{Delcamp:2020hzo}
C.~Delcamp and A.~Tilloy, ``{Computing the renormalization group flow of
  two-dimensional $\phi^4$ theory with tensor networks},''
  \href{http://dx.doi.org/10.1103/PhysRevResearch.2.033278}{{\em Phys. Rev.
  Res.} {\bfseries 2} no.~3, (2020) 033278},
  \href{http://arxiv.org/abs/2003.12993}{{\ttfamily arXiv:2003.12993
  [cond-mat.str-el]}}.

\bibitem{Meurice:2020pxc}
Y.~Meurice, R.~Sakai, and J.~Unmuth-Yockey, ``{Tensor lattice field theory for
  renormalization and quantum computing},''
  \href{http://dx.doi.org/10.1103/RevModPhys.94.025005}{{\em Rev. Mod. Phys.}
  {\bfseries 94} no.~2, (2022) 025005},
  \href{http://arxiv.org/abs/2010.06539}{{\ttfamily arXiv:2010.06539
  [hep-lat]}}.

\bibitem{Levin:2006jai}
M.~Levin and C.~P. Nave, ``{Tensor renormalization group approach to 2D
  classical lattice models},''
  \href{http://dx.doi.org/10.1103/PhysRevLett.99.120601}{{\em Phys. Rev. Lett.}
  {\bfseries 99} (2007) 120601},
  \href{http://arxiv.org/abs/cond-mat/0611687}{{\ttfamily
  arXiv:cond-mat/0611687}}.

\bibitem{NIST:DLMF0}
``{\it NIST Digital Library of Mathematical Functions}.''
  \url{https://dlmf.nist.gov/}, release 1.2.7 of 2026-06-15.
\newblock F.~W.~J. Olver, A.~B. {Olde Daalhuis}, D.~W. Lozier, B.~I. Schneider,
  R.~F. Boisvert, C.~W. Clark, B.~R. Miller, B.~V. Saunders, H.~S. Cohl, and
  M.~A. McClain, eds.

\bibitem{Bender:1969si}
C.~M. Bender and T.~T. Wu, ``{Anharmonic oscillator},''
  \href{http://dx.doi.org/10.1103/PhysRev.184.1231}{{\em Phys. Rev.} {\bfseries
  184} (1969) 1231--1260}.

\bibitem{Bender:1973rz}
C.~M. Bender and T.~T. Wu, ``{Anharmonic oscillator. 2: A Study of perturbation
  theory in large order},''
  \href{http://dx.doi.org/10.1103/PhysRevD.7.1620}{{\em Phys. Rev. D}
  {\bfseries 7} (1973) 1620--1636}.

\bibitem{Jentschura:2010zza}
U.~D. Jentschura, A.~Surzhykov, and J.~Zinn-Justin, ``{Multi-instantons and
  exact results. III: Unification of even and odd anharmonic oscillators},''
  \href{http://dx.doi.org/10.1016/j.aop.2010.01.002}{{\em Annals Phys.}
  {\bfseries 325} (2010) 1135--1172},
  \href{http://arxiv.org/abs/1001.3910}{{\ttfamily arXiv:1001.3910 [math-ph]}}.

\bibitem{Aguilar:1971ve}
J.~Aguilar and J.~M. Combes, ``{A class of analytic perturbations for one-body
  schroedinger hamiltonians},''
  \href{http://dx.doi.org/10.1007/BF01877510}{{\em Commun. Math. Phys.}
  {\bfseries 22} (1971) 269--279}.

\bibitem{Balslev:1971vb}
E.~Balslev and J.~M. Combes, ``{Spectral properties of many-body schroedinger
  operators with dilatation-analytic interactions},''
  \href{http://dx.doi.org/10.1007/BF01877511}{{\em Commun. Math. Phys.}
  {\bfseries 22} (1971) 280--294}.

\bibitem{Jones:2006qs}
H.~F. Jones and J.~Mateo, ``{An Equivalent Hermitian Hamiltonian for the
  non-Hermitian $-x^4$ potential},''
  \href{http://dx.doi.org/10.1103/PhysRevD.73.085002}{{\em Phys. Rev. D}
  {\bfseries 73} (2006) 085002},
  \href{http://arxiv.org/abs/quant-ph/0601188}{{\ttfamily
  arXiv:quant-ph/0601188}}.

\bibitem{Bender:2006wt}
C.~M. Bender, D.~C. Brody, J.-H. Chen, H.~F. Jones, K.~A. Milton, and M.~C.
  Ogilvie, ``{Equivalence of a Complex PT-Symmetric Quartic Hamiltonian and a
  Hermitian Quartic Hamiltonian with an Anomaly},''
  \href{http://dx.doi.org/10.1103/PhysRevD.74.025016}{{\em Phys. Rev. D}
  {\bfseries 74} (2006) 025016},
  \href{http://arxiv.org/abs/hep-th/0605066}{{\ttfamily arXiv:hep-th/0605066}}.

\end{thebibliography}
\end{document}